\documentclass{kapproc} % Computer Modern font calls
\usepackage{procps} 
\usepackage[dvips]{graphicx}
\upperandlowercase
\nochapequationnumber % will result in equation numbers that are (1)
\let\footnote\savefootnote
\let\footnotetext\savefootnotetext
\normallatexbib %will produce bibliography entries as shown in the LaTeX book

\begin{document}

\articletitle[]{Structure of Stars and Nuclei}

\author{J\"urgen Schaffner--Bielich}

\affil{Institut f\"ur Theoretische Physik, 
J. W. Goethe Universit\"at, 
D-60054 Frankfurt am Main, Germany}

\begin{abstract}
In these lectures, the properties of dense hadronic and quark matter and its
relation to compact stars will be discussed.  In a bottom--up approach
we start with nuclear and hypernuclear physics at low density and
extrapolate hadronic matter to large densities. The matching to the
quark matter phase is performed in a top--down approach starting at
asymptotically large densities. Implications for the mass--radius
relation of compact stars and the existence of a new family of solutions
will be outlined. 
\end{abstract}

\section{Introduction: Elementary Matter and Neutron Stars}

Neutron stars are created by supernovae type II explosions and are the
final endpoint of evolution of massive stars. The compact remnants of the
core collapse supernovae have masses in the range of 1--2 solar masses
and radii of the order of 10 km. The interior of neutron stars consists
of matter under extreme densities, several times the density of normal
nuclear matter, $n_0 = 3 \cdot 10^{14}$ g/cm$^3$. The study of neutron
stars has considerably advanced during the last years. With new
telescopes, ground-based and in satellites, one measures spectra of
supernova remnants like the crab nebula not only in the optical but also
in the x-ray (Chandra, XMM--Newton), in radio as well as in the infrared
band. The Hubble Space Telescope and the Chandra Telescope has even
published a movie of the crab nebula on--line! These movies demonstrate
how the rotating neutron star, the crab pulsar, pushes out energetic
wisps into the crab nebula in the equatorial plane as well as jets of
matter along the polar axis.

More than 1000 pulsars are known today. The masses of pulsars were
measured most precisely from a few binary neutron star systems
\cite{Thorsett99}, especially from the Hulse-Taylor pulsar with
$M=(1.4411\pm0.00035)M_\odot$. The shortest known rotation period so far
measured is 1.557 ms for the pulsar PSR 1937+21. 

Our knowledge about compact stars got a new twist by the discovery of
isolated, non-pulsating neutron stars. The first one seen and the closest
one known is RX J1856 \cite{Walter2001} being radio-quiet and with no
pulsations. The thermal spectrum indicates a temperature of 49 eV in the
optical band. The x-ray spectrum, however, shows a nice Planck curve
with a temperature of 60 eV as measured by the Chandra satellite
\cite{Drake02}. Most surprisingly, no spectral lines from elements in
the atmosphere of the neutron star have been found in the spectra!
These puzzles of the spectra of the isolated neutron star
are still not fully resolved (see \cite{Burwitz02} for a possible
explanations). The revised and improved parallax measurement of RX J1856
gives a distance of $D=117\pm 12$ parsec \cite{Walter02}. An ideal
black-body emitter would have a very small radius of $R_\infty = 4-8$ km
at that distance which is much smaller than the canonical value for a
neutron star of 10 km \cite{Drake02}. Corrections from an atmosphere
resulted in an apparent radius of $R_\infty = 15\pm 3$ km which would be
compatible with most of the modern neutron star models on the market
\cite{Walter02,Pons2002}.

The structure of neutron stars encompasses several distinctly different
zones when going from the surface to the centre (for a review see e.g.\ 
\cite{Haensel03}). First, there is a thin atmosphere up to about $10^4$
g/cm$^3$, mainly iron but could be also hydrogen or helium by accretion.
Then the outer crusts or the envelope begins which consists of free
electrons and nuclei forming a Coulomb lattice. The sequence of nuclei
starts with iron and then continues to neutron-rich nuclei stopping at
the neutron drip-line at about $10^{11}$ g/cm$^3$ \cite{BPS}. The inner
crusts consists of free neutrons in addition to nuclei and electrons
where the neutrons are in a superfluid state. The nuclei can now form
the pasta in the crust, various inhomogeneous phase structures as
bubbles (meat-balls), rods (spaghetti), and plates (lasagna) immersed in
the neutron and electron fluid. The sequence can then reverse, so that
the neutron fluid forms the geometrical structures immersed in a
background of extremely neutron-rich and superheavy nuclei
\cite{Negele73}.

At about half times normal nuclear matter density, the matter
distribution will be uniform, consisting of mainly neutrons with a small
admixture of protons and electrons, where the protons can now form a
superconductor. Still, the end of the crust is located far away from the
centre, as the crust is only a few hundred meters thick for massive
neutron stars!

The core of the neutron star consists of matter under extreme
densities. New forms of matter have been proposed to exist under these
condition in the very core of compact stars: Bose condensation of pions
or kaons, phase transitions to hyperon matter (hyperon star) and quark
matter (hybrid star). If strange quark matter is absolutely stable, then
the corresponding compact star is dubbed a strange star and the quark
phase extends all the way from the centre to the neutron-drip line, as
all free neutrons are swallowed by the true ground state of matter,
while nuclei are saved by virtue of the Coulomb barrier. 

Now let us start with a simple consideration: let us assume that the
high-density equation of state can be described by free (known)
particles. The stable baryons known in vacuum are the nucleons (n,p) and
hyperons (the $\Lambda$ and $\Sigma^{-,0,+}$ with one strange quark, the
$\Xi^{-,0}$ with two strange quarks, and the $\Omega^-$ with three
strange quarks). The masses of the hyperons increases with the number of
strange quarks. Besides those, there are spinless mesons which are
stable against strong interactions (charged pions, and kaons with one
anti-strange or one strange quark). They can form a Bose condensate. A
calculation of neutron star matter for a free gas of particles shows, that
first the negatively charged $\Sigma^-$ appears at about $4n_0$ and then
the $\Lambda$ at about $8n_0$ \cite{Ambart60}. The $\Sigma^-$ appears
before the $\Lambda$ despite the fact that it is slightly heavier,
because it is negatively charged. The presence of the $\Sigma^-$ can
then take over the r\^ole of the electron in making the overall matter
charge neutral, thereby lowering the Fermi energy of electrons and the
total energy of the system. One can compute that there will be no other
particles in the composition of a free gas of hadrons up to $20n_0$
which is well beyond the maximum density in the interior of a compact
star (besides that our approach will be not applicable anymore, as
hadrons are composite particles). The corresponding equation of state
will be even slightly softer than that of a a free neutron gas due to
the appearance of additional particles.  Hence, the maximum mass is the
one given by Oppenheimer and Volkoff for a free gas of neutrons which is
about $0.7M_\odot$ \cite{OV39}. That maximum mass is more than a factor
two smaller than required by the measurement of the mass of the
Hulse-Taylor pulsar of $1.44 M_\odot$!  Neutron stars can {\em not} be
described by a approximately free gas of particles, contrary to white
dwarfs. Interactions between hadrons are crucial for explaining such a
large neutron star mass.

\section{Nuclei with baryons: hypernuclei and strange hadronic matter}

As hyperons seem to appear in dense hadronic matter relevant for compact
stars, we study in the following the interactions between baryons of the
baryon octet (nucleons, $\Lambda$, $\Sigma$'s, and $\Xi$'s) in dense
matter. The nucleon-nucleon interaction is very well known from nuclear
properties (the general problem is, of course, how to extrapolate the
interaction to densities beyond normal nuclear matter density). For
hyperons the situation is far less clear but a lot of information can be
gathered from hypernuclear data and hyperonic atoms (for reviews see
e.g.\ \cite{Batty97,Gal02}). Most of the hyperon-nucleon and
hyperon-hyperon interaction known experimentally indicates an attractive
potential. 

The measured single particle levels of $\Lambda$ hypernuclei from mass
numbers of $A=3$ to 209 result in a potential of $U_\Lambda = - 30$ MeV
at $n_0$ (see e.g.\ \cite{Millener88,Rufa90}). It is well known, that the
$\Lambda$N interaction gets repulsive at densities of about $2n_0$ and
above \cite{Millener88,SMB97}, so this nonlinear behaviour with density has to
be taken into account for the modelling of neutron star matter.

For $\Sigma$'s, there is only one bound $\Sigma$ hypernucleus known,
$^4_\Sigma$He, which is, however, bound by isospin forces. $\Sigma^-$
atomic data indicates a repulsive potential at $n_0$ \cite{Mares95}
assuming again a nonlinear density dependence of the optical potential.
This finding is consistent with the absence of narrow $\Sigma$ hypernuclear
states (even in the continuum) for the reaction $^9$Be($K^-$,$\pi^-$) as
measured at BNL \cite{Bart99}. 

There are seven $\Xi$ hypernuclear events reported in the literature,
which are summarised in \cite{Dover83}. Correcting the $\Xi$ vacuum masses
for the older events, one arrives at a relativistic potential of about
$-28$ MeV at $n_0$ which is comparable to that of the $\Lambda$.
However, more recent indirect estimates of the $\Xi$N interaction by
final state interactions in the reaction ($K^-$,$K^+$) on $^{12}$C point
towards a reduced attraction of about $-14$ MeV \cite{Khaustov2000}.

Double $\Lambda$ hypernuclear states have been also seen experimentally
(see \cite{Gal02} for a summary). Besides the three older candidates,
there are two new measurements reported in 2001. Experiment E906 at BNL
finds indications of about 400 produced $_{\Lambda\Lambda}^{~~4}$H on a
$^9$Be target by the correlated emission of pions of the sequential weak
decays. Experiment E373 at KEK reconstructs the production of a
$_{\Lambda\Lambda}^{~~6}$He and its subsequent weak mesonic decays. There
are difficulties to reconcile the old and the new data in detail, but
the overall consensus is that the $\Lambda\Lambda$ interaction must be
attractive (see the discussion in \cite{Gal02}). 

Note, that all other hyperon-hyperon interactions, as $\Lambda\Sigma$,
$\Sigma\Sigma$, $\Lambda\Xi$, $\Sigma\Xi$, and $\Xi\Xi$ are essentially
unknown experimentally! 

The hyperon-hyperon interaction strength can be also probed by two
particle correlations of e.g.\ two $\Lambda$s in relativistic heavy-ion
collisions as it was measured recently by the NA49 collaboration
\cite{Blume02}. Unfortunately, the statistics does not allow for a
definite conclusion about the size of the s-wave scattering length of
two $\Lambda$s, although the data indicates a small scattering length.
As a lot of hyperons are produced in a single central heavy-ion
collision at relativistic bombarding energies, multiply strange nuclear
systems can be produced in the laboratory
\cite{Scha92,Scha93,Scha94,SMS00}. Indeed, the hypernuclei $^3_\Lambda$H
and $^4_\Lambda$H can be seen in the invariant mass spectra of the weak
decay products of Helium and a $\pi^-$ \cite{Finch99}.

Building up more and more hyperons in a ''nucleus'', the total binding
energy of the system increases, as new degrees of freedom are filled up
with a lower Fermi energy than nucleons. However, in the multi-hyperonic
medium new reaction channels are possible, as $\Xi^- + p \to 2 \Lambda$
and $\Xi^0 + n \to 2\Lambda$. These reactions are Pauli-blocked, as the
possible lowest lying $\Lambda$ hyperon states are filled up by
$\Lambda$'s. Hence, the whole multi-hyperonic nucleus can be stabilised
by virtue of the Pauli-blocking in the hyperon world! Relativistic shell
model calculations find that the binding energy can be $-13$ MeV/A or
even $-21$ MeV/A, depending on the strength of the hyperon-hyperon
interactions \cite{Scha93,Scha94,SG00}. Even purely hyperonic ''nuclei'',
consisting of only $\Lambda$, $\Xi^0$, and $\Xi^-$ hyperons can exist
with up to $-8$ MeV/A binding energy. More importantly, the charge of
these systems is negative, while the baryon number is positive --- 
indeed, a quite unusual nuclear property.

\section{Neutron Stars with baryons}

As outlined in the introduction, hyperons will appear within a free gas
of hadrons but interactions have to be taken into account to be compatible
with the observed neutron star masses. Our more or less profound
knowledge about the nucleon-nucleon, nucleon-hyperon, and
hyperon-hyperon, the latter being the least well known of all, can be
utilised to derive models for neutron star matter.  In many different
approaches emerges that hyperons, either the $\Lambda$ or the
$\Sigma^-$, appear in neutron star matter at about $2n_0$: in
relativistic mean-field models \cite{Glen85,Knorren95b,SM96}, in a
nonrelativistic potential model \cite{Balberg97}, in the quark-meson
coupling model \cite{Pal99}, in relativistic Hartree--Fock models
\cite{Huber98}, in Brueckner--Hartree--Fock calculations
\cite{Baldo00,Vidana00}, and within chiral effective Lagrangians
\cite{Hanauske00}. Nevertheless, as we will show in the following, the
details of the hyperon composition of neutron star matter is rather
sensitive to the chosen hyperon potentials. 
 
\begin{figure}
\centerline{\includegraphics[width=0.7\textwidth]{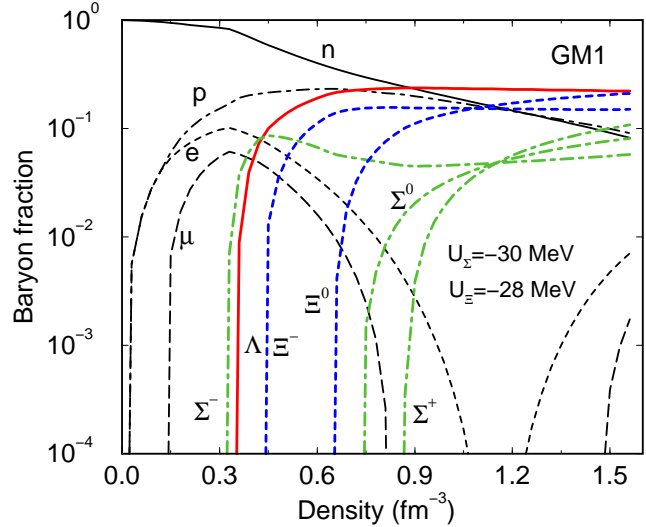}}
\caption{The composition of neutron star matter with hyperons within the
relativistic mean-field model using parameter set 1 of \cite{GM91} and
an attractive potential for the $\Sigma$ hyperons at $n_0$.}  
\label{fig:compold}
\end{figure}

Fig.~\ref{fig:compold} shows the composition of neutron star matter as a
function of density. At low densities, neutrons are dominating and there
is a rapid rise of the proton and electron fraction. At about $2n_0$,
first the $\Sigma^-$ appears closely followed by the $\Lambda$. The
electron fraction starts to drop at that density. At $3n_0$, the $\Xi^-$
is present in neutron star matter before the lighter hyperons $\Sigma^0$
and the $\Sigma^+$ appear by virtue of its negative charge. The hyperon
fraction considerably rises with increasing density. At $6n_0$ the
$\Lambda$ fraction even exceeds that of the neutron, being then the most
populated baryon in matter in $\beta$ equilibrium (which may more aptly
be called hyperon star matter). Neutron stars are giant
multi-hypernuclei \cite{Glen85}! Note, that at about $7n_0$, electrons
disappear and the isospin partners have equal fractions, as the isospin
potential is negligibly small at these densities. At that point, the
negatively charged $\Xi^-$ balances the positive charge of the protons!
At even larger densities, the matter distribution approaches about equal
fractions for all baryons and positrons can be present. The positrons
can not annihilate as there are no more electrons in the system left.
The overall effect of the presence of additional degrees of freedom in
the composition of a neutron star, be it hyperons or any other particle,
is to lower the pressure and therefore to soften the equation of state.
The maximum allowed mass for a neutron star will then decrease,
typically from say $2M_\odot$ for a purely nucleonic star to about
$1.5M_\odot$ for a hyperonic star depending on the parameter set used
for the nucleon--nucleon interaction.

\begin{figure}
\centerline{\includegraphics[width=0.7\textwidth]{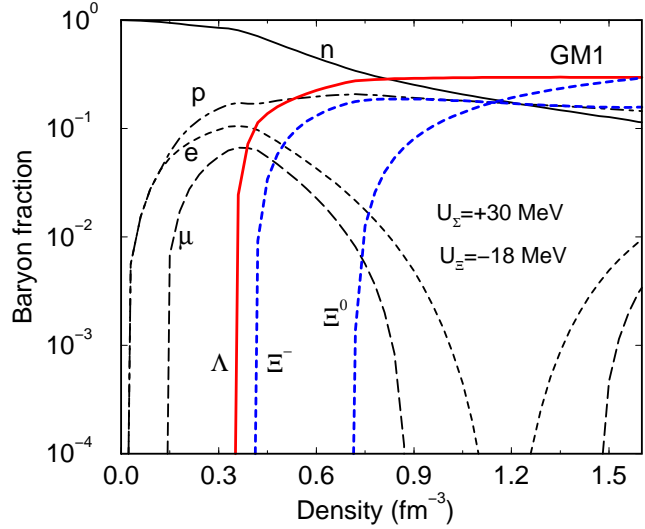}}
\caption{The composition of neutron star matter with hyperons, but now
with a repulsive potential for the $\Sigma$ hyperons at $n_0$.}  
\label{fig:compnew}
\end{figure}

Now what happens if one just change the hyperon potentials, in
particular switch the sign of the potential of the $\Sigma$ hyperons
from attraction to repulsion. Note that one adjusts the coupling
constant of the $\Sigma$ hyperons to the scalar meson slightly so as to
get a repulsive potential at $n_0$. All hyperon potentials are repulsive
above, say $2n_0$, as pointed out above, due to the nonlinear density
dependence of the baryon potentials. Furthermore, it is interesting to
know, that the baryon potentials are dominated by its coupling to the
vector mesons at large densities which are unaltered as their coupling
strengths are fixed solely by symmetry constraints (SU(6) symmetry).
Fig.~\ref{fig:compnew} depicts the matter composition for a repulsive
potential of the $\Sigma$ (at $n_0$) as indicated by $\Sigma^-$ atomic
data and a reduced attraction for the $\Xi$ as extracted from final
state interactions of $\Xi$s with $^{12}$C. Here, a relativistic
potential depth of $U_\Xi=-18$ MeV is taken, which is slightly larger
than the non-relativistic one. The $\Sigma$ hyperons are not present at
all up to $10n_0$, demonstrating how a small change in the parameters
can drastically shift its critical density. The only hyperons appearing
in matter are the $\Lambda$, $\Xi^-$, and the $\Xi^0$, in that order.
Nevertheless, the onset of the $\Xi^-$ population is shifted slightly
towards lower densities to $2.5n_0$ compared to Fig.~\ref{fig:compold}.
The $\Xi^-$ takes over the r\^ole of the $\Sigma^-$ to diminish the
fraction of electrons and its presence is even more favoured, if the
$\Sigma^-$ hyperons are absent. The $\Xi$ hyperons constitute now a
substantial fraction of the whole composition of hyperon star matter.
Note, that the electron fraction starts to drop even before the critical
density for the $\Xi^-$ is reached. Once the neutral $\Lambda$ fraction
rises, the system lowers its energy by replacing protons with
$\Lambda$s, thereby reducing again the Fermi energy of electrons which
are needed to balance the charge of the protons. One might think, that a
repulsive potential for the $\Xi$ would also remove them completely from
the matter distribution as the $\Sigma$s. However, this is not the case,
the $\Xi^-$ still appears, now at $3n_0$. The underlying reason is, that
the repulsive vector potential for the $\Xi$s is a factor of two smaller
than for the other hyperons and a factor three smaller than for nucleons
as demanded by SU(6) symmetry for the vector coupling constants. The
repulsive contribution to the overall potential, the vector potential,
has to dominate at large densities to avoid a collapse of matter. Hence,
the $\Xi$ is more favoured than every other baryon at large densities
due to its decreased repulsive potential and less sensitive to the
changes of the hyperon potential at $n_0$ than the $\Sigma$.

So far we have not discussed the interactions between hyperons. As
indicated by the scarce double hypernuclear data, the $\Lambda\Lambda$
interaction is attractive, maybe even more than the $\Lambda$--nucleon
interaction. There is definitely an additional attractive contribution
between two $\Lambda$s in nuclei, as the binding energy of two
$\Lambda$s is more than just twice the binding energy of a single
$\Lambda$. So additional potentials between hyperons have to be
considered. Here we follow the discussion in \cite{Scha94,SM96,Scha02}.
It is natural to assume that there is scalar and a vector potential only
between hyperons, as for nucleons and the nucleon--hyperon interaction.
Utilising SU(3) symmetry arguments for the meson exchanges, the
additional potentials appear to be just mediated by the missing members
of the scalar and vector meson nonet: a hidden strange scalar meson, the
$f_0(980)$, which we denote as $\sigma^*$ in the following, and the
hidden strange vector meson $\phi(1020)$. The vector coupling constant
to the $\phi$--meson can be fixed by SU(6) symmetry, so that all vector
coupling constants are given in terms of the nucleon one. In SU(6)
symmetry, there is ideal mixing between the singlet and octet states and
the vector coupling constants are proportional to the number of light or
strange quarks of the baryon, respectively. The conserved nonstrange
vector current is then proportional to the light quark number current,
while the hidden strange vector current is proportional to the strange
quark number current. The scalar meson coupling constant to the
$\sigma^*$ will be adjusted to the hyperon--hyperon interaction
strength. We assume that the $\sigma^*$ meson coupling constants scale
with the number of strange quarks, like for the $\phi$ meson coupling
constants. As a consequence of these arguments, the hyperon--hyperon
interaction between $\Xi$s should then be twice as strong as between
$\Lambda$s and $\Sigma$s. The overall strength of the hyperon--hyperon
interaction is then controlled by one parameter, the hidden strange
scalar coupling constant of the $\Lambda$ to $\sigma^*$.

\begin{figure}
\centerline{\includegraphics[width=0.7\textwidth]{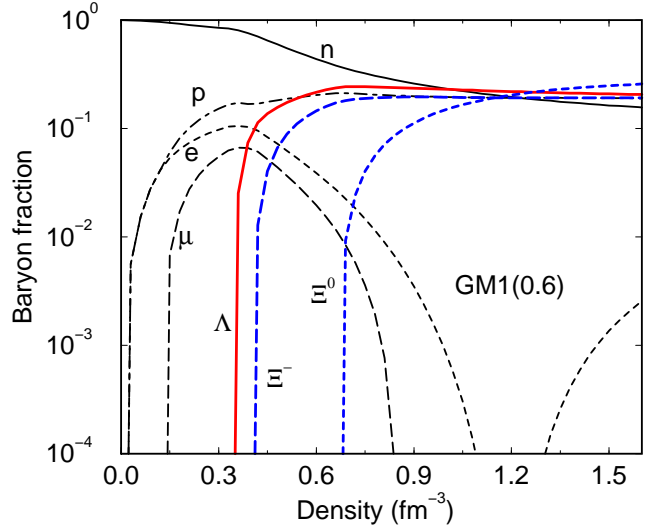}}
\caption{The composition of neutron star matter with hyperons including 
effects from an explicit hyperon--hyperon interaction
($g_{\sigma*\Lambda}/g_{\sigma N}=0.6$).}   
\label{fig:compyyweak}
\end{figure}

Fig. \ref{fig:compyyweak} shows the composition, if the additional
potentials for the hyperon--hyperon interaction are taken into account
with $g_{\sigma*\Lambda}/g_{\sigma N}=0.6$. As one can see, there is
not a substantial change in the composition of neutron star matter
compared to the case without explicit hyperon--hyperon
interactions. The appearance of hyperons is hardly modified. One
notices, however, that the hyperon fraction at large densities is
reduced, so that the hyperon fraction does not exceed the neutron
fraction anymore. The composition at large densities is now uniform,
there are about equal amounts of all baryons present (except for the
$\Sigma$ hyperons). The reduced fraction of hyperons at large
densities is due to the additional vector potential between hyperons
which dominates over the attractive hidden strange scalar potential at
large hyperon densities. The equal abundance of all baryons, except
for the $\Sigma$s, is due to the fact that the sum of vector
potentials of each baryon scales with the total number of quarks, not
with the number of light quarks.  The strange twist by including the
attractive hyperon--hyperon interaction is that the hyperon fractions
are reduced. The hyperon--hyperon interaction is attractive at normal
nuclear density and for a small hyperon population, but it turns
repulsive for a large hyperon density i.e.\ in the core of a neutron
star. The additional repulsions helps to stabilise the system at large
densities in the sense that the equation of state gets stiffer at
large densities.

\begin{figure}
\centerline{\includegraphics[width=0.7\textwidth]{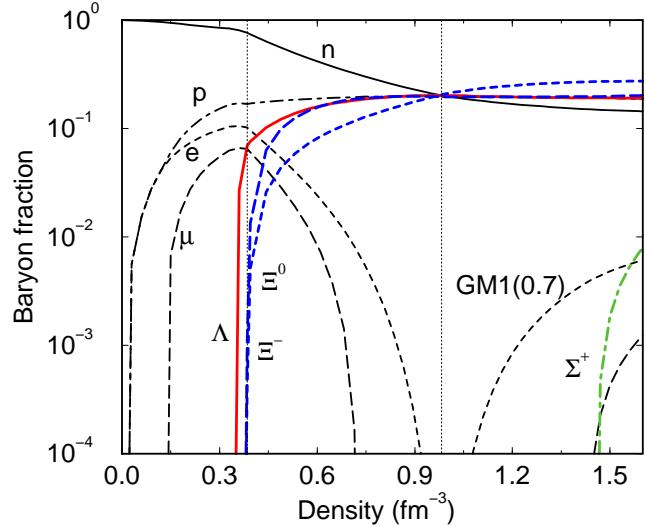}}
\caption{The composition of neutron star matter with hyperons, now for
a slightly increased hyperon--hyperon interaction
($g_{\sigma*\Lambda}/g_{\sigma N}=0.7$).  A first order phase
transition appears, the boundaries of the mixed phase are marked by
the vertical dotted lines.}
\label{fig:compyystrong}
\end{figure}

Now let us increase the strength of the hyperon--hyperon interaction
slightly from $g_{\sigma*\Lambda}/g_{\sigma N}=0.6$ to
$g_{\sigma*\Lambda}/g_{\sigma N}=0.7$. Fig.~\ref{fig:compyystrong}
depicts the composition of neutron star matter. Surprisingly, the
composition has drastically changed. The $\Lambda$ population sets in
slightly before the appearance of the $\Xi$ hyperons, which appear
together at about $2.5n_0$. At $6.5n_0$, the baryon fractions are
equal to each other and the electrons, as well as the muons, have
disappeared in the medium. A first order phase transition from nucleon
dominated to hyperon dominated hadronic matter appeared in the neutron
star matter at $2.5n_0$! Using the general Gibbs criterion to model
the phase transition (see e.g.\ \cite{Glen92}), the electron fraction
is continuous through the boundaries of the mixed phase. Negatively
charged bubbles of hyperon dominated matter form which are immersed in
slightly positively charged nucleon dominated matter. At the end of
the mixed phase at $6.5n_0$, the situation is reversed, now positively
charged bubbles of nucleon dominated matter is present in the
background of slightly negatively charged hyperon dominated matter. At
each stage, the criterion of total global (not local) charge
neutrality is fulfilled. The boundaries of the mixed phase are marked
by the vertical dotted lines in the figure. Note that the mixed phase
extends over a quite large region of density, thereby constituting a
substantial fraction of the total matter inside a neutron
star. Geometrical phases appear in the mixed phase, bubbles, rods and
slabs, like in the liquid--gas phase transition of nuclear matter in
the crust. The charged structures will form a Coulomb lattice, thereby
changing the liquid to a solid and modifying the transport properties
of the inner parts of a neutron star.

\begin{figure}
\vspace*{-0.8cm}
\includegraphics[width=0.65\textwidth]{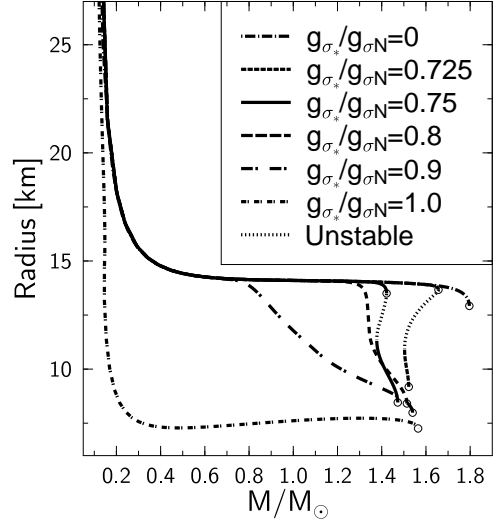}
\vspace*{-3.7cm}
\caption{The mass--radius relation of compact stars for different
interaction strengths of the hyperon--hyperon potential. For
increasing hyperon--hyperon attraction, the radius of the maximum mass
compact star decreases down to only $R=7-8$ km. In some intermediate
cases, an additional branch of stable solutions appears (taken from
\cite{Scha02}).} 
\label{fig:mryy}
\end{figure}

The presence of a first order phase transition in neutron star matter
has also an important impact on the global properties of compact stars
\cite{Scha02}. The mass--radius relation for compact stars is shown in
Fig.~\ref{fig:mryy} for different hyperon--hyperon interaction
strengths.  For a moderate hyperon--hyperon interaction strength,
$g_{\sigma*\Lambda}/g_{\sigma N}=0.725$ and 0.75, the maximum mass is
reduced while the corresponding radius is unaltered in the ordinary
neutron star branch. In addition, there appears a new branch of
solutions at smaller radii but similar (maximum) masses. The dotted
lines denote the unstable regions of the solution, where the compact
star is unstable with respect to radial oscillations. The dashed and
full lines then indicate that the solution is stable against radial
oscillations, as the mass increases with decreasing radii. The maximum
mass of this new branch of solution amounts to about $1.5M_\odot$ with
a radius of about only $7-8$ km. The compact stars representing this
new family of solutions are hypercompact stars. For comparison, the
maximum mass of the ordinary neutron star branch lies at similar
values while the radius is much larger, around 13 km. Note, that
hypercompact stars can have larger or lower masses than the maximum
mass of the ordinary neutron star branch.

Increasing the hyperon--hyperon interaction, the new family of solution
disappears and there is a continuous stable line in the mass--radius
diagram. The maximum mass compact star is, however, now always located
close to the hypercompact star branch, i.e.\ at about $1.5M_\odot$ with
a radius of about $7-8$ km. A new effect happens for the case
$g_{\sigma*\Lambda}/g_{\sigma N}=1.0$. For asymptotically large radii,
that means for low densities, the mass and radius is like in the other
cases. The masses of the compact stars stay, however, below $0.2M_\odot$
down to radii of $7-8$ km and increase with a rather constant radius up
to a maximum mass of about $1.6M_\odot$. The underlying equation of
state for this case is one which exhibits a vanishing pressure at a
finite value of the energy density, a feature characteristic for
absolutely stable matter, like absolutely stable strange quark matter
\cite{Witten84,Farhi84}. The corresponding compact stars are so called
selfbound, as they are stabilised by strongly attractive interactions,
contrary to ordinary compact stars which are hold together by the
gravitational force. The outermost layer of material of a selfbound
compact star consists of the same material as for the outer crust of an
ordinary neutron star: a lattice of nuclei in a sea of electrons. The
maximum density of that crust must be below the neutron drip density, as
free neutrons will be eaten up by the absolutely stable matter which is
more deeply bound than ordinary matter. Nuclei survive due to the
Coulomb barrier between them and the phase of absolutely stable matter.
There is a strong jump in baryon density, accompanied by a similar jump
in the electron density which constitutes the Coulomb barrier, of three
orders of magnitude at the phase boundary of nuclear matter and
absolutely stable matter.  As the crust material is the same for
selfbound stars and neutron stars, the mass-radius curves approach each
other for low central energy densities and large radii, where most of
the material is residing in the crust. The properties of selfbound
strange hadronic matter are similar to the ones of selfbound strange
quark matter \cite{Haensel86,Alcock86}.

\begin{figure}
\centerline{\includegraphics[width=0.9\textwidth]{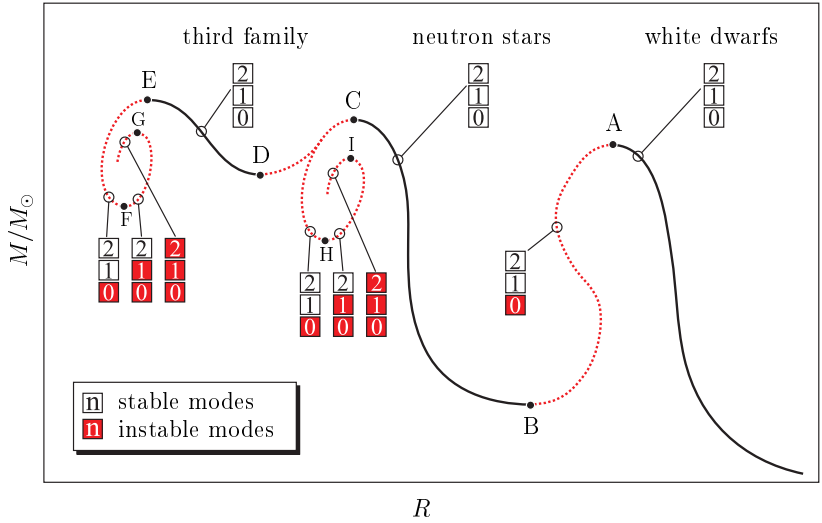}}
\caption{A sketch of the mass--radius relation of compact stars. The
curves with an increasing mass while lowering the radius are stable
solutions. A third solution can appear for radii below the ones for
ordinary neutron stars (taken from \cite{Schertler00}).}
\label{fig:mrtwins}
\end{figure}

Let us come back to the issue of the new branch of stable solutions.
Fig.~\ref{fig:mrtwins} demonstrates the general feature of the
mass--radius relation for compact stars. For white dwarfs, the mass
increases for decreasing radii up to the Chandrasekhar mass, the maximum
mass of white dwarfs. Beyond that limiting mass, the solutions are
unstable against radial oscillations, until the mass increases again for
decreasing radii. The family of neutron stars has been reached now,
which is again stable up to some maximum mass which depends sensitively
on the hadronic interactions. Beyond that maximum mass for neutron
stars, the mass--radius relation curves either in a (unstable) spiral
down to some fixed point or continues again to increase the mass with
lower radii.  The latter case is stable against radial oscillations and
marks the onset of a new family of stable solution of compact stars. The
masses of these new branch can have similar values as those of ordinary
neutron stars, so that they can be dubbed neutron star twins, but the
radii are considerably smaller. After the maximum mass of the third
family is reached, the unstable final spiral appears which marks the end
of Tolman--Oppenheimer--Volkoff (TOV) solutions for hadronic matter.
 
The existence of a new family of compact stars besides white dwarfs
and ordinary neutron stars stems from the presence of a strong first
order phase transition in the equation of state. The input for the TOV
equations is just the macroscopic relation between energy density and
pressure. The microscopical details are not relevant for the global
properties of compact stars. Hence, the first order phase transition
can be to anything, like to strange quark matter in hybrid stars
\cite{GK2000,Schertler00,FPS01}, to a pion condensed phase
\cite{Kaempfer81}, to a kaon condensed phase \cite{Banik01}, or, as
discussed above, to strange hadronic matter \cite{Scha02}. The
criterion for the existence of a new branch in the mass--radius
relation of compact stars has already been discussed by Gerlach in
1968 \cite{Gerlach68}. The new branch appears to be stable, if the
equation of state changes rapidly from a rather soft one to a much
harder one. Exactly this feature is present at the end of the mixed
phase of a strong first order phase transition, no matter what the
underlying reason for that strong phase transition is.

\section{Neutron stars with quarks}

In the previous section, we started with some interaction fixed to
properties of matter at low densities, that means at $n_0$, which is a
bottom--up approach for the description of dense, cold matter.  In
this section, we want to discuss the issue of the strong first order
phase transition in terms of a top--down approach. At sufficiently
large densities, the more appropriate description of dense matter will
be in terms of quarks, which are, according to our present
understanding of QCD, asymptotically free. 

For large temperatures and vanishing net baryon density, lattice gauge
simulations indicate, that the phase transition is first order for pure
gluon theory and likely to be a rapid crossover at about $T_c=170$ MeV
in full QCD including dynamical quarks \cite{Karsch04}. On the lattice,
it was seen that when the quark condensate drops at $T_c$
the Polyakov loop increases. The quark condensate is an order
parameter for the chiral phase transition, where the quarks are getting
massless for vanishing current quark masses for $T\geq T_\chi$. The
Polyakov loop is an order parameter for pure gluon theory and has a
nonvanishing value in the deconfined phase for $T\geq T_d$. Hence, the
lattice data demonstrates that $T_c=T_\chi =T_d$ at zero
quark-chemical potential. Why these two entirely different phase
transitions happen at the same temperature is presently not fully
understood. In principle, the chiral and the deconfinement phase
transitions can happen at entirely different scales in QCD.

\begin{figure}
\centerline{\includegraphics[width=0.75\textwidth]{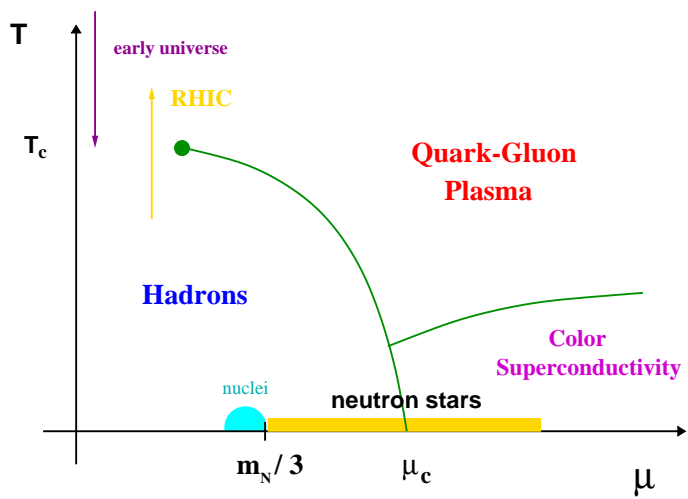}}
\caption{The phase diagram of QCD for finite temperature and quark
chemical potentials. There is a line of a first order phase transition
starting at zero temperature and a critical quark chemical potential
$\mu_c$ and stopping in a critical endpoint where the phase transition
is of second order (taken from \cite{Fraga03}).}
\label{fig:phasediagram}
\end{figure}

Neutron stars probe a very different region of the phase diagram of QCD
(see Fig.~\ref{fig:phasediagram}). Neutron star matter is cold on
nuclear scales and has an extremely large baryon density. At some
critical value of the quark chemical potential $\mu_c$, there is a first
order phase transition to colour superconducting quark matter
\cite{Barrois77,Bailin84,Alford98,Rapp98} which is based on theoretical
arguments (see \cite{KrishnaReview,TomReview,DirkReview,MarkReview} for
reviews about the many phases of QCD). Matter with colour
superconducting quarks occupies the phase diagram at large chemical
potentials and moderate temperatures. For large chemical potentials and
temperatures, there will be a quark--gluon plasma.  The line of first
order phase transition which started at $T=0$ and $\mu=\mu_c$ stops in a
critical point. At that point the phase transition is of second order,
beyond the point at larger temperatures the phase transition is a rapid
crossover. The precise location of that critical endpoint is not known
at present.

Recently, lattice simulations started to explore the phase diagram of
QCD at finite quark chemical potential attempting to find the critical
endpoint. The extension to finite $\mu$ is performed by reweighting
techniques or a Taylor expansion which will fail, of course, if $\mu/T$
gets large. So far, first lattice calculations find values of
about $\mu_{\rm end}=400-700$ MeV \cite{FK02,Karsch04}. In any case, it is
quite likely that there is a first order phase transition at finite
density and low temperature which is right in the region of interest for
neutron star matter.

We now study the following picture of cold and dense matter: at low
densities there are hadrons and at intermediate densities the
description of dense matter will be more appropriate in terms of massive
quarks. The transition from hadrons to massive quarks is assumed to be
smooth. Then there is a phase transition from massive quarks to massless
quarks at the chiral phase transition $\mu=\mu_\chi$. For
$\mu>\mu_\chi$, perturbative QCD is used as a model of the equation of
state of QCD and for $\mu<\mu_\chi$ a matching to the low density
equation of state has to be performed \cite{FPS01,MarkReview}. The
thermodynamic potential can be computed up to order $\alpha_s^2$ for
massless quarks from which all thermodynamic observables are derived
self-consistently (pressure, energy density, and baryon number density).
The renormalisation subtraction point needs to be fixed in the
perturbative treatment. It should be proportional to the overall scale
of the system and is chosen to be $\bar\Lambda/\mu=2,3$. One finds then,
that the resulting pressure starts at some finite value of the chemical
potential and approaches very slowly the Stefan--Boltzmann limit of a
free gas. Even at rather large densities, say of $n=30n_0$, there are
still sizable corrections to a simple free gas of quarks.

\begin{figure}
\centerline{\includegraphics[width=0.55\textwidth]{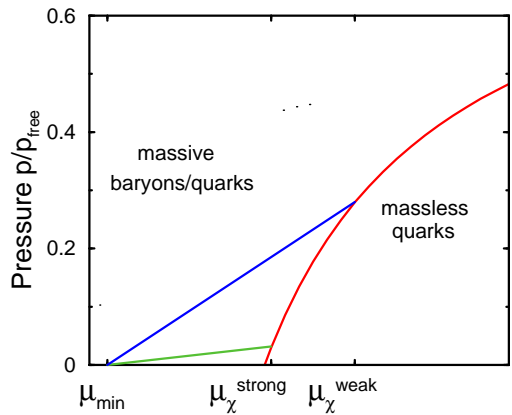}}
\caption{The matching of the low density equation of state to the high
density matter of massless quarks (taken from \cite{Fraga03}). Depending
on the rise of the pressure in the low density phase, the matching
results in a weak phase transition at $\mu_{\rm weak}$ (upper line) or
a strong phase transition at $\mu_{\rm strong}$ (lower line).}
\label{fig:matching}
\end{figure}

The matching to the low density equation of state can happen basically
in two ways: either the matching appears to be smoothly from the low
to the high density part or the matching incorporates a drastic change
of the slope of the pressure as a function of chemical
potential. Fig.~\ref{fig:matching} depicts the situation
schematically. If the pressure in the low density equation of state
increases rapidly with density, the matching will be smooth and the
phase transition is weakly first order. On the other hand, if the
pressure rises slowly with density the phase transition will be
strongly first order. Now the matching of the two equation of states
results in a significant change of the slope of the curves in
Fig.~\ref{fig:matching}. As the slope reflects just the number density,
the baryon density as well as the energy density will jump at the
matching point causing a strong first order phase transition. Which
situation is realized in nature, is not clear. Nonrelativistic model
calculations of asymmetric matter \cite{Akmal98} hint at that the
pressure in the low density regime stays small up to about $2n_0$,
being only a few percent of that of a free gas of quarks, so that a
strong first order phase transition is not excluded a priori.

\begin{figure}
\centerline{\includegraphics[width=0.64\textwidth]{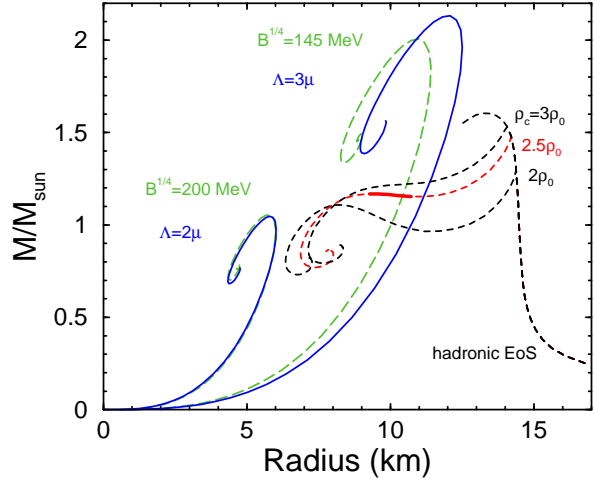}}
\caption{The mass radius relation of pure quark stars are shown by the
solid lines (perturbative QCD) and long--dashed lines (MIT bag
model). The mass--radius curves for a hadronic equation of state (EoS)
with a phase transition to quark matter are plotted with short--dashed
lines for different critical densities. A new stable solution appears
where the curve turns from a short--dashed to a solid line.}
\label{fig:mrquark}
\end{figure}

Fig.~\ref{fig:mrquark} shows the mass--radius relation for various
equation of states. The solid lines stand for pure quarks stars
calculated in perturbative QCD, the long--dashed ones are calculated
within the MIT bag model, a free gas of massless quarks only modified by
a constant bag pressure. The curves for quark stars start at zero mass
and radius as the pressure vanishes at a finite energy density --- no
hadronic mantle has been taken into account which would change the
mass--radius relation at small masses to large radii (see e.g.\ the
discussion of Fig.~\ref{fig:mryy}). The mass--radius relation for a
purely hadronic equation of state is depicted on the right side of the
plot. At some chosen critical density, the phase transition to quark
matter occurs. The corresponding hybrid stars have smaller radii. For
$n_c=2.5n_0$, a new family of stable solution appears where the
short--dashed line turns into a solid line. The matching is done for the
quark matter equation of state producing quark stars with a rather low
mass of only $1M_\odot$ and a corresponding radius of only 6 km.
Depending on the matching of the hadronic to the quark matter equation
of state, the new family of solution lies somewhere between the curves
for pure hadronic and pure quark stars, i.e.\ radii down to 6 km are in
principle possible for the new family of compact stars. We note in
passing, that there is a recent flurry of activity studying effects from
color superconducting quark matter on the mass-radius relation of
compact stars
\cite{Alford03,Baldo03,Banik03,Blaschke03,Shovkovy03,Ruster04,Buballa03}.

It is clear from the previous discussion, that the studies of compact
stars is far from being completed. Further investigations of the
nonperturbative treatments of the quark phase and how it matches to
the low density equation of state are needed to elucidate the possible
existence of a new family of compact stars.  The recent data from
present x-ray satellites, Chandra and XMM Newton, and the Hubble Space
telescope have considerably advanced our knowledge of the properties of
compact stars with more surprising news to come. Future telescopes,
like the x-ray satellite XEUS and the Next Generation Space Telescope, are
poised to finally pin down the mass--radius relation of compact stars
and to determine the equation of state of cold and dense,
strongly interacting matter. Excitingly, this means also that the
possible existence of hypercompact stars might be confirmed within
the coming years!

\begin{acknowledgments}

It is a pleasure to thank Eduardo Fraga, Norman Glendenning, Matthias
Hanauske, Rob Pisarski, and Igor Mishustin with whom I collaborated on
the topics addressed in these lectures and Mark Alford for comments. I
am indebted to Walter Greiner and Horst St\"ocker for their continuous
support which made this work possible. I especially thank Walter Greiner
for inviting me to give these lectures. 

\end{acknowledgments}

% Bibliography made with BibTeX:
%Use use the bibliographic style 'amsunsrt.bst'
\bibliographystyle{prsty}
\chapbblname{proc_kemer_jsb} % \chapbblname{<name of .bbl file>}
\chapbibliography{all,literat} % \chapbibliography{ <name of .bib file>}

\end{document}